\documentclass[twocolumn,showpacs,preprintnumbers,amsmath,amssymb,prd]{revtex4}
\usepackage{graphicx}
\usepackage[colorlinks=true]{hyperref}

\newcommand{\comment}[1]{}

\newcommand{\lr}[1]{ \left( #1 \right) }
\newcommand{\lrs}[1]{ \left[ #1 \right] }

\newcommand{\tr}{ {\rm Tr} \: }

\newcommand{\const}{ {\rm const \:}  }

\newcommand{\expa}[1]{ \exp{\left( #1 \right)} }

\begin{document}
\sloppy

\preprint{ITEP-LAT/2007-08}

%%% a in Fermies, on all plots!!!

\title{Center vortices as rigid strings}
\author{P. V. Buividovich}
\email{buividovich@tut.by}
\affiliation{Belarusian State University, 220080 Belarus, Minsk, Nezalezhnasti av. 4}
\author{M. I. Polikarpov}
\email{polykarp@itep.ru}
\affiliation{ITEP, 117218 Russia, Moscow, B. Cheremushkinskaya str. 25}
\date{May 25, 2007}
\begin{abstract}
 It is shown that the action associated with center vortices in $SU(2)$ lattice gauge theory is strongly correlated with extrinsic and internal curvatures of the vortex surface and that this correlation persists in the continuum limit. Thus a good approximation for the effective vortex action is the action of rigid strings, which can reproduce some of the observed geometric properties of center vortices. It is conjectured that rigidity may be induced by some fields localized on vortices, and a model-independent test of localization is performed. Monopoles detected in the Abelian projection are discussed as natural candidates for such two-dimensional fields.
\end{abstract}
\maketitle

\section{Introduction}
\label{sec:Introduction}

 Yang-Mills theory is often believed to be equivalent to some string theory, however, up to now there is no way to detect thin strings behind physical chromoelectric string of finite thickness, which gives rise to linear potential between test colour charges and which is usually seen in numerical simulations \cite{Bali:95:1, Gubarev:07:1}. On the other hand, closed magnetic strings, or center vortices, can be directly detected \cite{DelDebbio:98} and seem to be thin \cite{Polikarpov:03:1}. A model-independent argument in favor of physically thin center vortices in continuum pure Yang-Mills theory was proposed recently in \cite{Buividovich:07:2}. It is known that the full QCD string tension is reproduced with sufficiently good precision if one considers only the contribution due to topologically nontrivial winding of center vortices and Wilson loop \cite{DelDebbio:98, Bornyakov:01:1}. Moreover, in \cite{Polikarpov:03:1, DelDebbio:98} it was demonstrated that the area of center vortices in $SU\lr{2}$ LGT scales in physical units of length. These facts imply that center vortices are not just lattice artifacts, but rather correspond to some physically significant objects. Center vortices are usually detected using the maximal center gauge and are seen as closed self-avoiding surfaces, which occupy only a small fraction of lattice plaquettes \cite{Polikarpov:03:1, DelDebbio:98}. Typically one observes a large percolating vortex, which extends through the whole lattice, and a number of small satellite vortices, whose size typically does not exceed several lattice spacings \cite{Polikarpov:06:1}.

\begin{figure}[ht]
  \includegraphics[width=5cm, angle=-90]{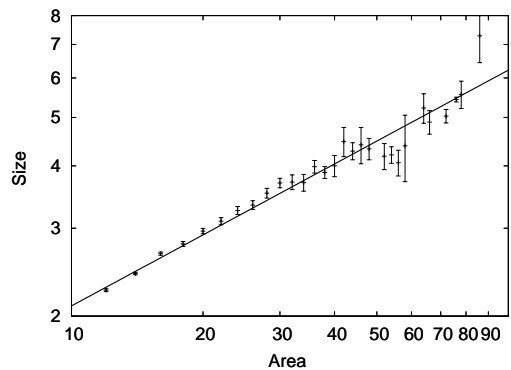}\\
  \caption{Average size of center vortices as the function of their area ($\beta = 2.60$, $28^{4}$ lattice)}
  \label{fig:size_vs_area}
\end{figure}

 As the total physical area and size of center vortices remain finite in the continuum limit, there should be also a continuous description of vortex geometry, which is characterized by some finite Hausdorf dimension $d_{H}$. Percolating vortex by definition has Hausdorf dimension equal to $4$. In order to define the dimensionality of small satellite vortices in $SU\lr{2}$ LGT, their size $L$ was measured as the function of their area $S$. The size of the vortex was defined as the maximal distance between points which belong to the vortex. Average size of small vortices as the function of their area (in lattice units) is plotted on Fig. \ref{fig:size_vs_area}. Fit of the form $L = \const \cdot S^{1/d}$ (solid curve on Fig. \ref{fig:size_vs_area}) gives $d = 1.9 \pm 0.1$. For small vortex areas ($S \lesssim 30$)  this number simply reflects self-avoiding of vortex surfaces, but for larger values of area this fit indicates that small vortices tend to be smooth surfaces.

 How can one describe the properties of such surfaces? A necessary condition for physical scaling of area of random surfaces is the cancelation between their entropy, which for self-avoiding surfaces grows linearly with the area of  surface in lattice units, and the bare string tension, which therefore should be finite in lattice units \cite{Ambjorn:94:1}. A remarkable result of \cite{Polikarpov:03:1} is that the excess of action associated with center vortices is indeed proportional to the area of vortices in lattice units. However, for the simplest model of random surfaces with Nambu-Goto action such naive balance between action and entropy does not lead to physical surfaces because of branched polymer problem \cite{Ambjorn:94:1}, therefore one has to consider some more complicated model.

 A well-studied model which can probably describe smooth surfaces even in four and three dimensions is the model of rigid strings \cite{Polyakov:86:1, Kleinert:86:1, Kleinert:88:1, Ambjorn:93:1, Ambjorn:94:1}. The model was first analyzed in \cite{Polyakov:86:1, Kleinert:86:1}, where it was shown that depending on the form of the $\beta$-function of the model either the branched polymer phase or the phase of smooth surfaces can be observed. Namely, if the $\beta$-function has no nontrivial fixed points, the model reduces to the usual Nambu-Goto string flawed by branched polymer problem, but if nontrivial fixed point exists, in the vicinity of this point the model should describe smooth surfaces with finite Hausdorf dimension \cite{Polyakov:86:1}. Numerical simulations indeed confirmed the existence of the phase of smooth surfaces in this model \cite{Ambjorn:93:1, Koibuchi:05:1, Ambjorn:94:1}. Nontrivial scaling in the vicinity of the phase transition was observed in \cite{Ambjorn:93:1}, but in more recent simulations of a related model of tethered surfaces first-order phase transition was found \cite{Koibuchi:05:1}. However, even if the model of rigid strings has no continuum limit, it can still serve as an effective model, for which UV cutoff is set by some other fields in the theory. It is known, for instance, that rigidity terms arise in the effective string action after integrating over worldsheet fermions \cite{Sedrakyan:86:1, Wiegmann:89:1, Ambjorn:94:1} or some four-dimensional massive fields coupled to the string worldsheet \cite{Kleinert:88:1, Akhmedov:96:1, Orland:94:1}.

 The aim of this paper is to fit observed properties of center vortices in the model of rigid strings. This model was first conjectured to describe center vortices in \cite{Engelhardt:00:1, Polikarpov:03:1}. In this paper the correlation between the excess of action associated with vortices and the geometry of vortex surfaces will be studied systematically. In the section \ref{sec:EffActFit} it will be shown that the effective vortex action can be indeed approximated by the action of rigid strings and that the coupling constants before extrinsic and internal curvatures are finite in the continuum limit. In the section \ref{sec:TwoDimFields} this action is discussed as an effective action induced by some two-dimensional fields localized on the surfaces of vortices and a model-independent check of localization is performed. Monopoles in the Abelian projection of the theory \cite{Ambjorn:00:1, Polikarpov:03:1, Polikarpov:05:1} are then discussed as appropriate candidates for such localized fields in the section \ref{sec:MonopAct}. The dependence of the bare string tension on the lattice spacing can be explained if one takes into account the contribution of percolating monopole cluster. Interaction between monopoles, which can be approximated by the Yukawa potential with physical mass \cite{Suzuki:98:1}, can also partially account for surface rigidity \cite{Kleinert:88:1, Akhmedov:96:1, Orland:94:1}. Finally, implications and possible extensions of obtained results are discussed.

\section{Effective action as the function of vortex geometry}
\label{sec:EffActFit}

\begin{figure}[ht]
  \includegraphics[width=5cm, angle=-90]{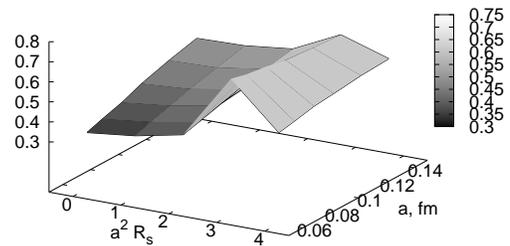}\\
  \caption{Average excess of action per site as the function of extrinsic curvature and lattice spacing}
  \label{fig:act_vs_excurv_spacing}
\end{figure}

\begin{figure}[ht]
  \includegraphics[width=5cm, angle=-90]{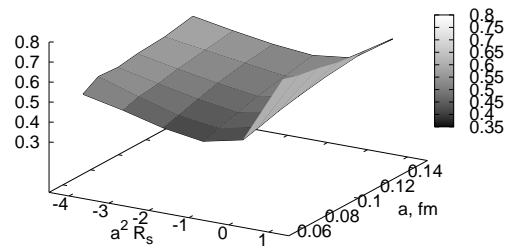}\\
  \caption{Average excess of action per site as the function of internal curvature and lattice spacing}
  \label{fig:act_vs_intcurv_spacing}
\end{figure}

 In order to measure the correlation between geometric properties of vortex surface and the action density center vortices were detected in $SU\lr{2}$ LGT by imposing the direct maximal central gauge (DMC) \cite{Polikarpov:03:1}. Simulated annealing procedure was used to locate true minima. In this work the same set of lattice configurations as in \cite{Polikarpov:03:1} was used. The lattices used were of the size $28^{4}$ for $\beta = 2.60$ and $\beta = 2.55$, $24^{4}$ for $\beta = 2.50$, $\beta = 2.45$ and $\beta = 2.40$ and $16^{4}$ for $\beta = 2.35$. Lattice spacing was fixed by setting the value of QCD string tension to $\sqrt{\sigma} = 440$ MeV.

 Local geometry of vortex surfaces was characterized by the two simplest local invariants -- internal and extrinsic curvatures. Internal curvature in lattice units for hypercubic lattice was defined as $a^{2} R_{s} = 4 - n_{s}$, where $n_{s}$ is the number of neighbors of the site $s$ and $a$ is the lattice spacing \cite{Ambjorn:94:1}. Extrinsic curvature for smooth surfaces can be written as $K = \Delta x^{\mu} \Delta x^{\mu}$, where $\Delta f = \frac{1}{\sqrt{g}} \partial_{a} \lr{ \sqrt{g} g^{a b} \partial_{b} f } $ is the two-dimensional Laplacian on the surface and $g_{a b} = \partial_{a} x^{\mu} \partial_{b} x^{\mu}$ is the induced metric on the surface. In order to define extrinsic curvature on hypercubic lattice two-dimensional lattice Laplacian on the surface was defined as $a^{2} \Delta f_{s} = n_{s} f_{s} - \sum \limits_{s'} f_{s'}$, where $s'$ are the sites adjacent to $s$. Extrinsic curvature in lattice units is then $a^{2} K_{s} = \Delta x^{\mu}_{s} \Delta x^{\mu}_{s}$ \cite{Ambjorn:94:1}. The excess of action associated with some lattice site on the vortex was defined by averaging the excess of action over all vortex plaquettes which contain this site. The excess of action on plaquette was defined as $\Delta S_{p} =  \beta \lr{1 - \frac{1}{2} \tr U_{p} } - \langle \beta \lr{1 - \frac{1}{2} \tr U_{p} } \rangle  = \frac{\beta}{2} \lr{ \langle \tr U_{p} \rangle - \tr U_{p}}$. Average excess of action per site as the function of lattice spacing and extrinsic and internal curvatures (in lattice units) is plotted on Figs. \ref{fig:act_vs_excurv_spacing} and \ref{fig:act_vs_intcurv_spacing} respectively.

\begin{figure}
  \includegraphics[width=5cm, angle=-90]{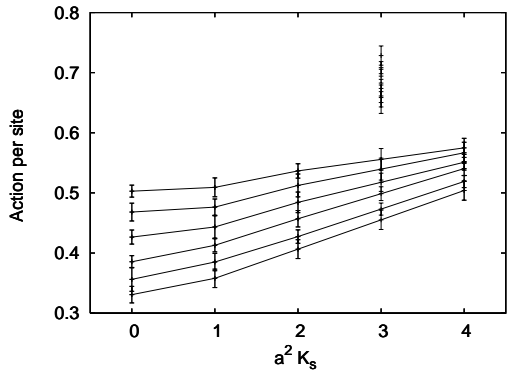}\\
  \caption{Excess of action per site as the function of extrinsic curvature at different lattice spacings. Solid lines are plotted using the value at $a^{2} K_{s} = 3$ averaged between $a^{2} K_{s} = 2$ and $a^{2} K_{s} = 4$}
  \label{fig:act_vs_excurv_slice}
\end{figure}

 It can be seen that the excess of action increases with both extrinsic and internal curvature, therefore the action should in general depend on both internal and extrinsic curvatures. Standard dimensional analysis shows that only the terms linear in extrinsic and internal curvature are relevant in the continuum limit. It should be also noted that because center vortices are defined on hypercubic lattice, typical values of curvature diverge as $a^{-2}$ in the continuum limit ($a^{2} K_{s}$ and $a^{2} R_{s}$ are finite integer numbers, therefore $K_{s} \sim a^{-2}$ and $R_{s} \sim a^{-2}$). A simple estimation shows that the number of points on the surface with $K \neq 0$ or $R \neq 0$ scales to zero as $a^{2}$, which ensures the smoothness of surfaces and the finiteness of the total contribution of bent plaquettes to the physical action. Finite values of $K_{s}$ or $R_{s}$ for physical surfaces near the continuum limit can be in principle defined by averaging the curvature over physically small regions, whose area is nevertheless very large in lattice units.

 A peculiar feature of the dependence of action excess on extrinsic curvature is the strong peak near $a^{2} K_{s} = 3$. Such value of extrinsic curvature corresponds to an edge where three plaquettes join. It was found that when this peak is neglected, the resulting function appears to be almost linear in $a^{2} K_{s}$ (Fig. \ref{fig:act_vs_excurv_slice}). Presumably the peak at $a^{2} K_{s} = 3$ is a lattice artifact and only reflects the anisotropy of hypercubic lattice. In order to obtain better fits the excess of action at $a^{2} K_{s} = 3$ was replaced by the average between $a^{2} K_{s} = 2$ and $a^{2} K_{s} = 4$ (solid line on Fig. \ref{fig:act_vs_excurv_slice}), which yields almost linear dependence. Bare string tension in lattice units and the coefficient before extrinsic curvature were obtained as the intercept and the slope of the linear fits of the data on Fig. \ref{fig:act_vs_excurv_slice}. One could in principle try to fit the excess of action per plaquette  by some polynomial in $a^{2} K_{s}$, but anyway dimensional analysis shows that only the term linear in $a^{2} K_{s}$ would survive in the continuum limit. Nevertheless, the fitting method affects numerical values of the coefficient $\kappa\lr{a}$ before extrinsic curvature. In general, increasing the number of degrees of freedom in fitting functions changes the value and the uncertainty of this coefficient, however, these values agree within error range when extrapolated to the continuum limit. For instance, in order to check the stability of fits the data plotted on Fig. \ref{fig:act_vs_excurv_slice} were fitted by first and second-order polynomials in $a^{2} K_{s}$. The values of bare string tension $\sigma_{0}\lr{a}$ (constant term in the fits) obtained from both fits agree with a very good precision. At finite lattice spacings discrepancies in the values of $\kappa$ are larger, but extrapolation to the continuum limit gives consistent values $\kappa = 0.065 \pm 0.006$ (linear fit) and $\kappa = 0.048 \pm 0.014$ (quadratic fit). The coefficient before $\lr{a^{2} K_{s}}^{2}$ contains very large errors and is close to zero. This coefficient can only be important in the continuum limit if it contains divergences of order $a^{-2}$, which is not likely.

 In order to extract the term linear in internal curvature from the data plotted on Fig. \ref{fig:act_vs_intcurv_spacing} the excess of action was fitted by a third-order polynomial in $a^{2} R_{s} = 4 - n_{s}$. The coefficient before $a^{2} R_{s}$ in this polynomial was assumed to be the coefficient before internal curvature in the physical action.

 Finally, after omitting all terms which become irrelevant in the continuum limit, for sufficiently small lattice spacings $a$ one can write the action associated with center vortices in the following form:
\begin{eqnarray}
\label{ActionFit}
 W\lrs{S} = \int \limits_{S} d^{2} \xi \sqrt{g} \lr{ \sigma_{0}\lr{a} \Lambda_{UV}^{2} + \gamma \lr{a} R + \kappa \lr{a} K }
\end{eqnarray}
where $\sqrt{g} = \sqrt{\det{g_{ab}}}$ is the invariant element of area on the surface, $g_{a b} = \frac{\partial X^{\mu}}{\partial \xi^{a}} \frac{\partial X^{\mu}}{\partial \xi^{b}}$ is the induced metric on the surface and $\Lambda_{UV} = a^{-1}$ is the UV cutoff scale of the theory.

\begin{figure}[ht]
  \includegraphics[width=5cm, angle=-90]{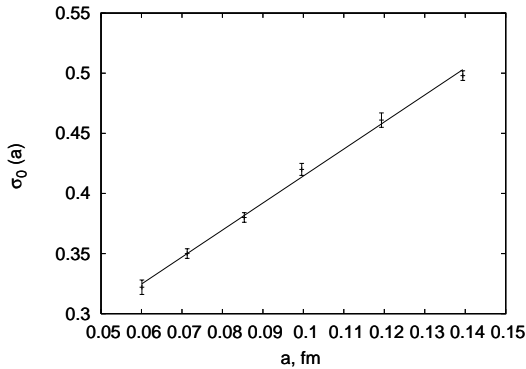}\\
  \caption{Bare string tension in lattice units $\sigma_{0} \lr{a}$ }
  \label{fig:sigma_vs_spacing}
\end{figure}

\begin{figure}[ht]
  \includegraphics[width=5cm, angle=-90]{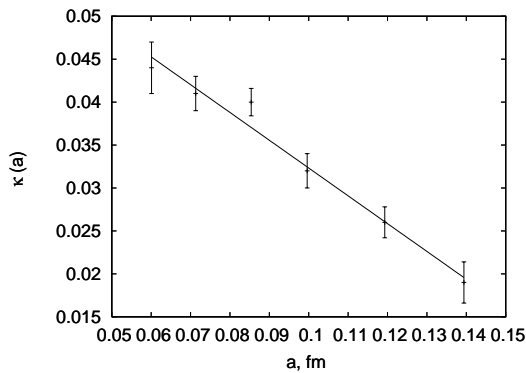}\\
  \caption{Coupling constant before extrinsic curvature $\kappa \lr{a}$}
  \label{fig:kappa_vs_spacing}
\end{figure}

\begin{figure}[ht]
  \includegraphics[width=5cm, angle=-90]{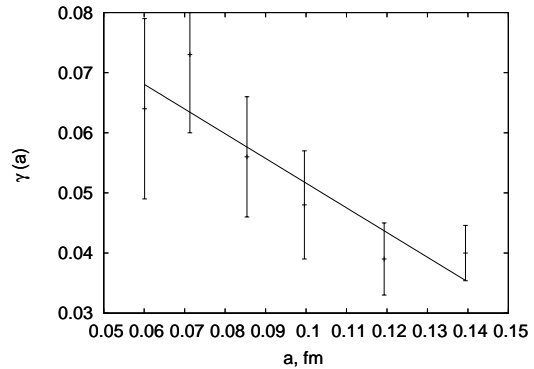}\\
  \caption{Coupling constant before internal curvature $\gamma \lr{a}$}
  \label{fig:gamma_vs_spacing}
\end{figure}

 The coefficients $\sigma_{0}\lr{a}$, $\gamma\lr{a}$ and $\kappa \lr{a}$ as the functions of lattice spacing are plotted on Figs. \ref{fig:sigma_vs_spacing}, \ref{fig:gamma_vs_spacing} and \ref{fig:kappa_vs_spacing} respectively. Extrapolation to continuum limit gives the following values:
\begin{eqnarray}
\label{ContinuumLimitParams}
\sigma_{0}\lr{0} = 0.192 \pm 0.006 \nonumber \\
\kappa \lr{0} = 0.066 \pm 0.003 \nonumber \\
\gamma\lr{0} = 0.08 \pm 0.02
\end{eqnarray}

 Thus coefficients $\kappa$ and $\gamma$ are finite in the continuum limit and therefore the dependence on internal and extrinsic curvatures is physical. Quadratic divergence in the bare string tension is crucial for the existence of smooth physical surfaces, as explained above, and should be compensated by a similar divergence in the entropy of random surfaces. It is interesting to note that the value of bare string tension in lattice units (\ref{ContinuumLimitParams}), which is obtained after taking curvature into account, is smaller than the value of action excess $ \langle \Delta S_{p} \rangle = 0.540 \pm 0.004$ obtained in \cite{Polikarpov:03:1}. The fact that after proper redistribution of action excess among operators with appropriate dimensions the string tension is strongly decreased indicates that the terms with extrinsic curvature play an important role in the dynamics of center vortices.

 The action (\ref{ActionFit}) corresponds to the model of rigid strings \cite{Polyakov:86:1, Kleinert:86:1, Ambjorn:94:1}. While the surface entropy is canceled by the divergent bare string tension, branched polymer problem is circumvented due to the third term. It is interesting to note that as the consequence of the Gauss-Bonnet identity $\int \limits_{S} d^{2} \xi \sqrt{g} R = 2 \pi \chi\lrs{S}$ ($\chi\lrs{S}$ is the Euler characteristic of the surface), the coefficient before internal curvature is proportional to the string coupling constant, which  is therefore also finite in the continuum limit.

\section{Fields localized on center vortices}
\label{sec:TwoDimFields}

 From the point of view of field theory quadratically divergent term $\int \limits_{S} d^{2} \xi \sqrt{g} \sigma_{0}\lr{a} \Lambda_{UV}^{2}$ in the vortex effective action indicates that some fields are localized on center vortices and become effectively two-dimensional. Quadratic divergence in the effective action (\ref{ActionFit}) corresponds to vacuum oscillations of these two-dimensional fields \cite{Zakharov:06:1, Zakharov:03:1}. Localization on center vortices was also directly observed in lattice simulations -- for instance, it was found that in the maximal Abelian gauge almost all trajectories of Abelian monopoles belong to center vortices \cite{Polikarpov:05:1}. There also exist classical string solutions where monopoles are localized on string worldsheets \cite{Gorsky:05:1, Chernodub:05:1}. In \cite{Greensite:06:1, Kovalenko:07:1} it was shown numerically that eigenfunctions of covariant four-dimensional Laplace and Dirac operators are also localized near center vortices.

 If only two-dimensional fields and their interactions are responsible for the excess of action on the vortices, total excess of action on the vortex $W\lrs{S}$ should be treated as the effective string action $W_{eff}\lrs{S}$ obtained by integrating over the fields at fixed geometry:
\begin{eqnarray}
\label{TwoDimEffAct}
\expa{ - W_{eff} \lrs{S} } = \int \mathcal{D} \phi
\expa{ - \int \limits_{S} d^{2} \xi \sqrt{g} \: L\lrs{\phi}  }
\end{eqnarray}
where $\mathcal{D} \phi$ is the covariantly defined path integral measure and $L\lrs{\phi}$ is the Lagrangian of the field $\phi$. Such effective action will necessarily contain quadratic divergence of the form $\int d^{2}\xi \sqrt{g} \: \Lambda^{2}_{UV}$ due to vacuum oscillations of these two-dimensional fields plus some geometry-dependent terms. For instance, if the fields $\phi$ are four-dimensional Dirac fermions, the effective action (\ref{TwoDimEffAct}) depends on the extrinsic curvature of the surface  \cite{Sedrakyan:86:1, Wiegmann:89:1, Ambjorn:94:1}.

 In order to check whether some two-dimensional fields propagate along the vortex or not, one can consider the points on the vortex which are very distant in terms of internal geometry on the vortex but are very close in four-dimensional space. If the relevant fields propagate only along the vortex, correlation between excess of action in such points should be much less than between plaquettes separated by the same distance along the vortex. In order to check this numerically center vortices were represented as graphs, each node of the graph corresponding to some lattice plaquette. The correlation between action densities was measured for the points which are separated by only one lattice spacing in four-dimensional space but no less than 6 spacings along the vortex ($d_{4} < 2$, $d_{2} \ge 6$). The standard breadth first search algorithm for unweighted graphs was used to measure the distances on vortex. In order to reduce anisotropy the sites which belong to the vortex were linked not only along lattice links, but also along the diagonals of plaquettes. As the distance measured was used only for lower-bound estimates of distances, there was no necessity to use much more time-consuming search algorithms for weighted graphs such as Dijkstra algorithm. For comparison the correlation between neighboring vortex plaquettes ($d_{4} < 2$, $d_{2} < 2$) was also measured. Correlation between plaquette variables $\tr U_{p}$ and $\tr U_{p'}$ was characterized by the correlation coefficient $\rho \lrs{\tr U_{p} , \tr U_{p'}}$:
\begin{eqnarray}
\label{CorrCoeff}
\rho\lrs{\tr U_{p},\tr U_{p'}} = \frac{\langle \tr U_{p} \: \tr U_{p'} \rangle - \langle \tr U_{p} \rangle^{2}}{ \langle \lr{\tr U_{p}}^{2} \rangle - \langle \tr U_{p} \rangle^{2} }
\end{eqnarray}

\begin{figure}[ht]
  \includegraphics[width=5cm, angle=-90]{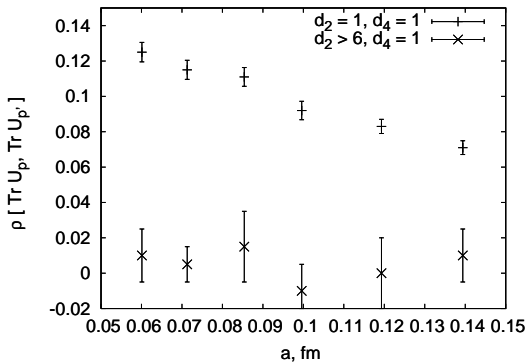}\\
  \caption{Correlation between neighbouring plaquettes with  $d_{4}<2$, $d_{2} \ge 6$ and $d_{4} < 2$, $d_{2} < 2$}
  \label{fig:plaq_corr}
\end{figure}

 The results of these measurements are plotted on Fig. \ref{fig:plaq_corr}. It can be seen that the correlation in four-dimensional space is notably smaller than along the surface of the vortex, and therefore the fields which are responsible for the excess of action on vortices are more likely to propagate along their surfaces. Unfortunately due to insufficient statistics it was not possible to measure the correlation lengths which correspond to propagation along the vortex and in four-dimensional space. The latter can be estimated as the inverse mass of the lowest glueball ($ \sim 1.5$ GeV), which is comparable with the inverse lattice spacing. A measurement which is somewhat similar to the one described above was performed in \cite{Polikarpov:03:1}, where the average excess of action on plaquettes very close to center vortices was shown to be zero.

\section{Abelian monopoles and center vortices}
\label{sec:MonopAct}

 Up to now the only way to see directly the content of the conjectured two-dimensional field theory is to impose maximal Abelian gauge and to look at the trajectories of Abelian monopoles \cite{Ambjorn:00:1, Polikarpov:03:2, Zakharov:06:1, Zakharov:03:1}, which populate densely the surfaces of center vortices. In real simulations about $95 \%$ of monopole trajectories belong to center vortices \cite{Ambjorn:00:1, Polikarpov:05:1}. It is natural to conjecture that two-dimensional field theory living on vortex surfaces should describe these monopoles upon first quantization.

 Another nontrivial fact which supports the statement about the role of monopoles in the dynamics of center vortices is the dependence of bare string tension in lattice units on lattice spacing (see Fig. \ref{fig:sigma_vs_spacing}), which can be approximated by a linear function $\sigma_{0}\lr{a} = A + B a $ with good precision. Besides quadratic divergence, this also yields $a^{-1} \sim \Lambda_{UV}$ divergence in the bare string tension in physical units: $a^{-2} \sigma_{0}\lr{a} = a^{-2} A + a^{-1} B $, where $A$ is given by (\ref{ContinuumLimitParams}) and $B = 2.2 \pm 0.1 fm^{-1}$. Such divergence usually corresponds to self-energy of one-dimensional objects \cite{Ambjorn:94:1, Polikarpov:03:1, Zakharov:06:1, Zakharov:03:1}. If the density of one-dimensional objects per unit of vortex area $\rho_{1D}$ scales in physical units and is constant in the continuum limit, while the bare mass of these objects is UV divergent and is close to the critical value $m_{bare} = a^{-1} \ln 4$ \cite{Ambjorn:94:1}, for $\rho_{1D}$ one then obtains the following estimation $\rho_{1D} = \frac{B}{\ln 4} = 1.5(8) fm^{-1}$. Taking into account that the density of vortices is $\rho_{v} = 24 fm^{-2}$, it is easy to estimate the density of one-dimensional objects in four-dimensional space as $\rho_{1D}' = 37.(9) fm^{-3}$, which is in a good agreement with the density of percolating Abelian monopoles $\rho_{m} = 31.(1) fm^{-3}$ \cite{Polikarpov:03:2}. The difference may be explained by incomplete detection of Abelian monopoles and by curved geometry of vortex surfaces.

 Geometric properties of monopole trajectories in $SU(2)$ LGT were studied in \cite{Polikarpov:03:2}. It was found that the properties of monopole trajectories at hadronic scale are not described by random walks with $d_{H} = 2$, as in the case of free scalar particles, but rather by smooth random walks. Namely, the correlation between the directions of tangent vectors to monopole paths is characterized by a correlation length $l_{c}^{-1} \approx 300$ MeV, which remains finite in the continuum limit. Smooth random walks are in the same universality class as the trajectories of spinning particles \cite{Ambjorn:94:1}, therefore the fields associated with monopoles living on center vortices are more likely to have nonzero spin. The simplest physical model which leads to smooth random walks is the propagation of massive Dirac fermions in Euclidean space \cite{Plyushchay:90:1, Ambjorn:94:1}. If the monopoles are assumed to be Dirac fermions, the mass of the fermions can be roughly estimated as $\sim 1.5$ GeV from the measurements of monopole current correlators in \cite{Polikarpov:03:2}. A general smooth random walk corresponds to random walk of the tangent vector on the three-sphere $S^{3}$ and therefore includes components of all spins, but it is not clear how such model can be incorporated into the models of random surfaces. On the other hand, random surfaces with four-dimensional fermions are well studied. It is known that the effective action of fermionic strings includes rigidity terms \cite{Sedrakyan:86:1, Wiegmann:89:1, Ambjorn:00:1}. As in the model of fermionic strings worldsheet fermions are massless Dirac fermions, it is natural to conjecture that the effective action induced by massive worldsheet fermions also induces string rigidity. In general, one can expect that for massive fermions the effective action besides the local terms of the form (\ref{ActionFit}) should also contain nonlocal terms of the form $\int d^{2} \xi_{1} \sqrt{g\lr{\xi_{1}}} \int d^{2} \xi_{2} \sqrt{g\lr{\xi_{2}}} K\lr{\xi_{1}} \Delta^{-1}\lr{\xi_{1}, \xi_{2}} K\lr{\xi_{2}}$, where $\Delta^{-1}\lr{\xi_{1}, \xi_{2}}$ is the kernel of the inverse Laplace operator on the surface of the vortex. Unfortunately, at presently available lattices it is very difficult to trace such terms in the effective action.

 The effective Lagrangian governing the dynamics of monopoles was obtained in \cite{Suzuki:98:1} using the inverse Monte-Carlo method. It was found that the effective monopole action, besides the usual kinetic term, contains Yukawa interaction with physical mass as well as four-point and six-point interactions. Higher-order interaction terms were found to be very small. Taking these results together, a reasonable conjecture is that at hadronic scale monopoles behave as massive Dirac fermions living on center vortices. Yukawa-type interaction may be induced if these fermions are coupled to some four-dimensional massive field. Coupling to massive four-dimensional fields also leads to rigidity terms in the effective action \cite{Kleinert:88:1, Orland:94:1, Akhmedov:96:1}, and thus such interaction can be also absorbed in the effective action (\ref{ActionFit}).

\section{Conclusions}
\label{sec:Discussion}

 In this paper the relation between local vortex geometry and the action density was studied. Direct measurements of action imply that a reasonable approximation for the effective action of center vortices is the action of rigid strings \cite{Polyakov:86:1, Kleinert:86:1}. This action can reproduce observed smoothness of vortex surfaces and presumably the physical scaling of vortex area. The latter possibility depends crucially on the form of nonperturbative $\beta$-function of the model of rigid strings and at present time can only be checked numerically \cite{Ambjorn:93:1}. Unfortunately the values of the parameters $\sigma_{0}$, $\kappa$ and $\gamma$ (\ref{ContinuumLimitParams}), obtained by extrapolation to the continuum limit can not be compared with the corresponding critical values obtained from independent simulations, because most numerical investigations of the model of rigid strings dealt with three-dimensional case \cite{Ambjorn:93:1, Koibuchi:05:1}. As the existence of the continuum limit of the model of rigid strings has not been proven exactly, it is not clear whether the effective vortex action (\ref{ActionFit}) can be used at all values of lattice spacing.

 It turns out that a large fraction of the action associated with center vortices comes from rigidity term, therefore the dependence of action on local vortex geometry should be crucial for the dynamics of vortices. In \cite{Polikarpov:03:1} it was conjectured that this dependence arises due to Abelian monopoles localized on vortices. Considerations of the section \ref{sec:MonopAct} support this conjecture, although this problem requires more accurate analytic treatment. For instance, it could be extremely interesting to construct two-dimensional field theory which describes monopoles localized on center vortices. Presumably such theory should be fermionic.

 An important property of center vortices which is not captured by the action (\ref{ActionFit}) is the existence of a single percolating vortex. In the case of random walks percolating trajectory corresponds to "condensate" which emerges due to tachyonic instability of perturbative vacuum. In order to describe condensation one should use the concepts of Euclidean interacting quantum field theory instead of random walks which describe the states of only one particle. Condensate then corresponds to nonzero background field, as in the Higgs model (for a related discussion see, e.g, \cite{Zakharov:06:1, Zakharov:03:1}). It is not clear whether this picture remains valid for the theory of random surfaces, because required nonperturbative apparatus of string field theory is almost not developed.

\begin{acknowledgments}
The authors are grateful to E. T. Akhmedov, F. V. Gubarev, A. S. Gorsky, and especially to V. I. Zakharov, for illuminating discussions and critical remarks and to F. V. Gubarev, A. V. Kovalenko and P. Yu. Boyko for lattice configurations and source codes. P. V. Buividovich is grateful to all members of the ITEP lattice group for their kind hospitality. M. I. Polikarpov was partially supported by grants RFBR-05-02-16306a, RFBR-0402-16079a, RFBR-0602-04010-NNIOa and EU Integrated Infrastructure Initiative Hadron Physics (I3HP) under contract RII3-CT-2004-506078.
\end{acknowledgments}

%\bibliography{MyBibliography}
%\bibliographystyle{apsrev}

\end{document}